\documentclass[11pt]{article}
\usepackage[a4paper,margin=1in]{geometry}
\usepackage[T1]{fontenc}
\usepackage[utf8]{inputenc}
\usepackage{lmodern}
\usepackage{microtype}
\usepackage{graphicx}
\usepackage{amsmath,amssymb}
\usepackage{booktabs,tabularx,multirow}
\usepackage{algorithm}
\usepackage{algpseudocode}
\usepackage[authoryear,round]{natbib}
\usepackage{url}
\usepackage[hidelinks]{hyperref}
\usepackage{caption}
\usepackage{float}
\graphicspath{{figures/}}
\title{CCP-DP: Cross-cluster context-guided phase dynamic programming for DNA storage sequence reconstruction}
\author{Zixin Zhang$^{1}$, Guanjin Qu$^{1}$, and Huaming Wu$^{1,2,*}$\\[4pt]
\small $^{1}$Center for Applied Mathematics, KL-AAGDM, Tianjin University, China\\
\small $^{2}$State Key Laboratory of Synthetic Biology, Tianjin University, China\\
\small $^{*}$Corresponding author: \href{mailto:whming@tju.edu.cn}{whming@tju.edu.cn}}
\date{}
\begin{document}
\maketitle
\begin{abstract}
\textbf{Motivation:} DNA data storage offers high information density and long-term stability, but accurate sequence reconstruction remains challenging under limited sequencing coverage and strong insertion, deletion, and substitution (IDS) noise. Existing reconstruction methods mainly rely on local evidence within individual read clusters. When sequencing coverage is low or indel-induced phase shifts become pronounced, such local evidence may be insufficient to distinguish closely competing candidate sequences.

\textbf{Results:} We present Cross-Cluster Context-Guided Phase Dynamic Programming (CCP-DP), which uses dataset-level cross-cluster information to complement cluster-local evidence and selectively refines ambiguous provisional consensuses. Under the main benchmark settings, CCP-DP achieved exact reconstruction rates of 95.36\%--99.98\% and outperformed existing methods on multiple datasets. At $5\times$ coverage, CCP-DP improved exact reconstruction by 6.2--17.5 percentage points over the stronger of BBS and CPL across the tested IDS conditions, while maintaining low reconstruction error and computational overhead. Overall, CCP-DP introduces cross-cluster shared information as a complementary source of evidence for DNA sequence reconstruction under challenging conditions.

\textbf{Availability:} Source code is available at \url{https://github.com/ZixinZ3899/CCP-DP}.
\end{abstract}
\medskip
\section{Introduction}

\begin{figure*}[!t]
  \centering
  \vspace{-0.1in}\includegraphics[width=0.97\textwidth]{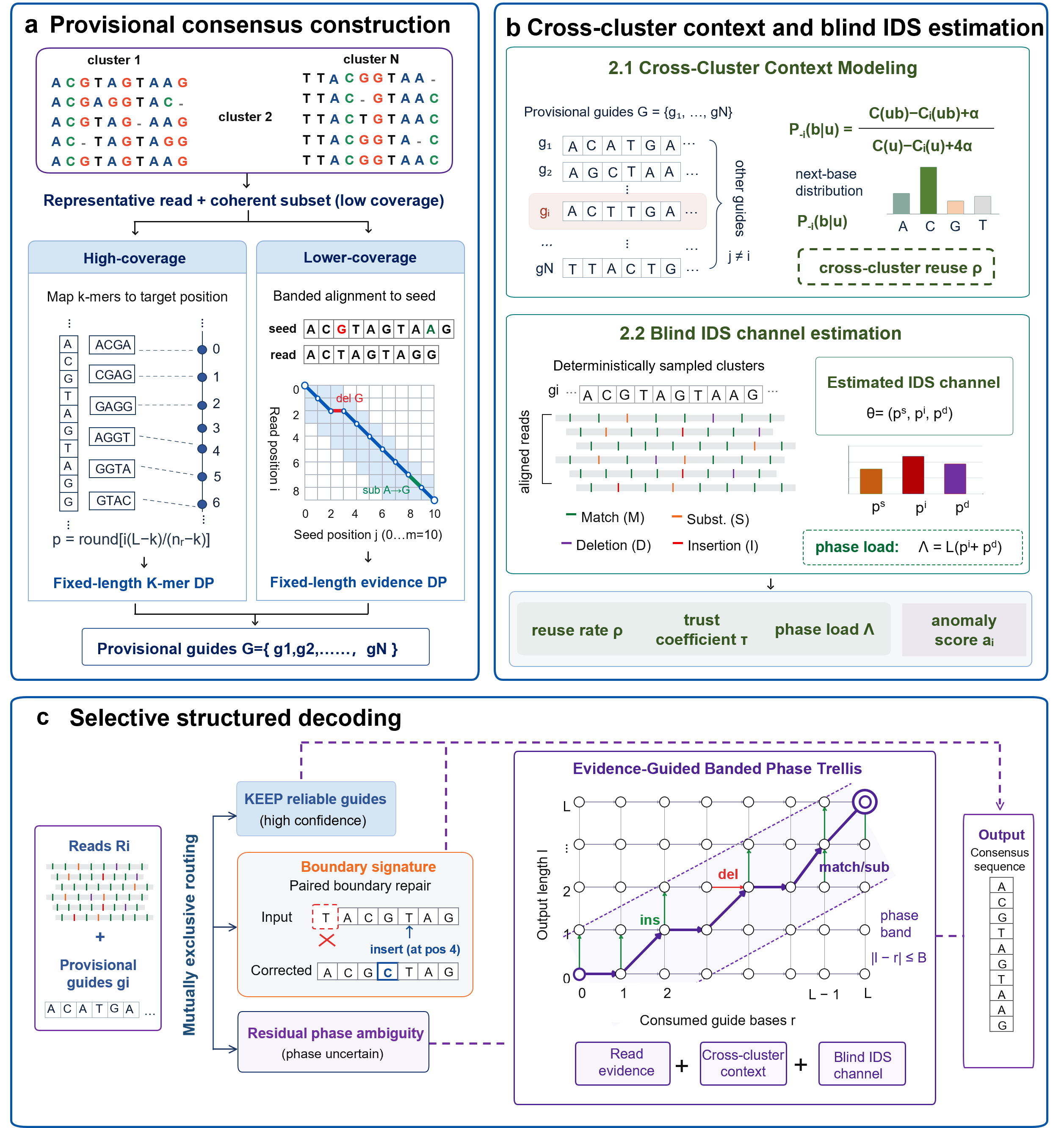}
  \caption{Overview of the proposed CCP-DP framework. The method first
  constructs a provisional consensus for each read cluster, then estimates cross-cluster contextual statistics and the blind IDS channel from the dataset. Based on the resulting reliability information, CCP-DP selectively retains the provisional consensus, applies structured boundary correction, or performs evidence-guided phase dynamic programming to obtain the final reconstruction.}
  \label{fig:method}
\end{figure*}

With the rapid growth of global data, conventional storage media are facing
increasing pressure in storage capacity and long-term preservation
\citep{ozturk2026dna,qu2025dna}. DNA has emerged as a promising storage medium owing to its extremely high storage density and long-term stability \citep{rebimbas2026dna}. During DNA synthesis and sequencing, insertion, deletion, and substitution (IDS) errors are introduced into the reads, and their characteristics depend on the underlying experimental platform
\citep{gimpel2023digital,yan2025dna,zhang2025gc}.  Third-generation sequencing technologies represented by Oxford Nanopore sequencing have provided a new option for DNA storage readout \citep{volkel2025nanopore,zhang2026spacecraft}.
However, nanopore sequencing generally has higher IDS error rates and more complex error patterns, particularly insertions and deletions. Therefore, accurate and efficient sequence reconstruction under complex IDS errors and limited sequencing coverage remains an important challenge in DNA data recovery.

A variety of methods have been developed for DNA sequence reconstruction.
MUSCLE-based approaches perform multiple sequence alignment within each read
cluster and derive a consensus from the aligned reads \citep{antkowiak2020low}. TrellisBMA uses trellis-based inference to account for sequence misalignment
caused by insertions and deletions \citep{srinivasavaradhan2021trellis},
whereas ITR iteratively refines candidate sequences according to the
consistency among reads \citep{sabary2024reconstruction}. CPL estimates
first-order conditional probabilities from pairwise read alignments and
searches for an optimal reconstruction path in the resulting graph
\citep{barlev2025scalable}.  BBS further combines high-order $k$-mer
statistics with bidirectional beam search to improve
reconstruction efficiency \citep{gu2025efficient}. More recently, deep-learning-based methods have also been
introduced. RobuSeqNet models read reliability with a Conformer-based
architecture \citep{qin2024robust}, DNAFormer combines DNN-based
reconstruction with confidence-guided refinement
\citep{barlev2025scalable}, and TReconLM formulates trace reconstruction as
a next-token prediction problem using a Transformer
\citep{weindel2026trace}.

Despite these advances, reconstruction decisions are still driven mainly by
evidence within the target cluster~\citep{qu2022clover}. When the sequencing coverage is low, or when frequent insertion and deletion errors cause substantial sequence misalignment, the limited local
evidence within a single cluster may be insufficient to distinguish
between several similar candidate sequences reliably. In fact, clusters within the
same DNA dataset usually undergo the same synthesis and sequencing pipeline,
and are therefore affected by similar channel error mechanisms
\citep{wang2023modelling}. In addition, similar local sequence contexts may
recur across different clusters. Such cross-cluster information can provide
additional evidence for clusters with insufficient local observations, but
has not been fully exploited by existing reconstruction methods.

Based on this observation, we propose Cross-Cluster Context-Guided Phase
Dynamic Programming (CCP-DP), which exploits dataset-level cross-cluster
contextual information and IDS statistics to provide additional evidence for
the reconstruction of difficult clusters.

\section{Materials and methods}

In this study, we propose Cross-Cluster Context-Guided Phase Dynamic
Programming (CCP-DP) for efficient reconstruction of original DNA
sequences from noisy-read clusters affected by insertion, deletion, and
substitution (IDS) errors. The overall framework is illustrated in
Fig.~\ref{fig:method}. CCP-DP first constructs a target-length
provisional consensus for each cluster using its local reads. It then
extracts shared cross-cluster sequence context and dataset-level IDS
channel statistics from all provisional consensuses and observed reads,
and uses this information as complementary evidence beyond cluster-local
reconstruction. For clusters that remain uncertain, CCP-DP further
integrates local and cross-cluster evidence for structured phase-aware
inference, while clusters with sufficient support retain their provisional
consensuses.

\subsection{Provisional Consensus Construction}
\label{sec:local_consensus}

The first stage of CCP-DP constructs a provisional consensus of target
length $L$ for each read cluster using only cluster-local observations.
A representative read is selected as the local reference, and a reliable
subset of reads is retained according to their consistency with the
reference to reduce the influence of anomalous reads.

Because the reliability and redundancy of local evidence depend on cluster
coverage, CCP-DP uses two complementary local reconstruction strategies. For
high-coverage clusters, read $k$-mers are projected onto a common target
coordinate system, and candidate $k$-mers are connected through
$(k-1)$-overlap constraints. Their support is accumulated within a local
neighborhood to tolerate small indel-induced positional shifts, and dynamic
programming selects the maximum-support target-length phase chain.
For lower-coverage clusters, selected reads are band-aligned to the
representative read. The alignments are summarized as base/deletion evidence
at reference positions and insertion evidence between adjacent positions.
These local event scores are then jointly decoded by a length-constrained
dynamic program to produce a provisional consensus of length $L$.
Detailed formulations of both branches are provided in Supplementary Methods
and Supplementary Figs.~S1--S2.

\subsection{Cross-Cluster Context and IDS Estimation}

After provisional consensus construction, CCP-DP further exploits shared
information across clusters within the same dataset. Because these clusters
typically undergo the same synthesis and sequencing process, similar local
sequence contexts may recur across clusters, and their reads may exhibit
similar insertion, deletion, and substitution (IDS) error characteristics.
Based on these shared properties, CCP-DP constructs a cross-cluster
sequence-context model and a blind IDS channel model from the current dataset,
providing additional statistical evidence for subsequent refinement.

\subsubsection{Cross-Cluster Context Model}

Let $G=\{g_i\}_{i=1}^{N}$ denote the set of provisional consensus sequences. To quantify local-context reuse across clusters, let $d(u)$ denote the number of
provisional consensus sequences containing $k$-mer $u$, counting repeated
occurrences within the same consensus only once. The cross-cluster reuse rate is defined as
\begin{equation}
\rho
=
\frac{
\sum_{i,t}
\mathbb{I}\!\left[d(u_{i,t})\geq 2\right]
}{
\sum_{i,t}1
},
\label{eq:cross_reuse}
\end{equation}
where $u_{i,t}$ denotes the $k$-mer starting at position $t$ in provisional
consensus $g_i$. A larger \(\rho\) indicates broader cross-cluster context reuse.

We next construct a leave-one-cluster-out (LOO) model from the provisional
consensus set $G$. Let $u$ denote a context of length $k$, and let $b$ be
its succeeding nucleotide. For 
target cluster $i$, the LOO conditional probability is defined as
\begin{equation}
P_{-i}(b\mid u)
=
\frac{
C(u,b)-C_i(u,b)+\alpha
}{
C(u)-C_i(u)+4\alpha
},
\label{eq:loo_context}
\end{equation}
where $C(\cdot)$ denotes the accumulated count over $G$, $C_i(\cdot)$
denotes the contribution of $g_i$, and $\alpha=0.5$ is the additive
smoothing constant. By subtracting the contribution of
$g_i$ from the global statistics, $P_{-i}(b\mid u)$ is determined solely by
contextual information from the remaining provisional consensus sequences.
The reverse context is computed in the same manner and combined with the
forward context to provide cross-cluster contextual support.

Because the usefulness of cross-cluster context varies across datasets,
CCP-DP introduces a dataset-level trust coefficient $\tau \in [0,1]$ to
control the contribution of Cross-LOO evidence. The coefficient reflects
the reliability of cross-cluster context in the current dataset by shrinking
the Cross-LOO distribution toward the uniform distribution:
\[
\widetilde{P}_{-i}(b\mid u;\tau)
=(1-\tau)\frac{1}{4}+\tau P_{-i}(b\mid u).
\]
The value of $\tau$ is selected using separate calibration and validation
transitions derived from the provisional guides, without using ground-truth
sequences. Cross-cluster evidence is retained only when it shows consistent
predictive gain under this internal validation; otherwise, its contribution
is suppressed. The resulting $\tau$ controls the Cross-LOO weight in
Phase-DP. Detailed candidate settings and acceptance criteria are provided
in the Supplementary Methods.

In addition to assessing whether cross-cluster context is reliable at the
dataset level, CCP-DP further evaluates whether each provisional consensus
still contains pronounced local reconstruction inconsistencies. These
inconsistencies are assessed jointly from the Cross-LOO context and the
within-cluster read evidence, and are aggregated into a cluster-level anomaly
score $a_i$. A larger $a_i$ indicates a greater concentration of unresolved
reconstruction ambiguity within the cluster. Accordingly, $\rho$, $\tau$,
and $a_i$ characterize cross-cluster reuse, calibrated context reliability,
and cluster-level uncertainty, respectively, and are used jointly to determine
the scope of subsequent structured decoding. Detailed position-level scoring
and aggregation rules are provided in the Supplementary Methods.

\subsubsection{Blind IDS Channel Model}

In addition to cross-cluster context, CCP-DP estimates a dataset-specific IDS
profile from deterministically sampled clusters. Reads are band-aligned to their corresponding provisional consensuses, after which
alignment outliers are discarded. Match, substitution, insertion, and
deletion events are then accumulated as $M$, $S$, $I$, and $D$,
respectively. Let $B=M+S+D$. The
corresponding IDS parameters are estimated as
\begin{equation}
\hat{\boldsymbol{\theta}}_{\rm IDS}
=
\left(
\hat{p}_s,\hat{p}_i,\hat{p}_d
\right)
=
\left(
\frac{S}{B},
\frac{I}{B},
\frac{D}{B}
\right),
\label{eq:ids_channel}
\end{equation}
where $\hat{p}_s$, $\hat{p}_i$, and $\hat{p}_d$ denote the effective
substitution, insertion, and deletion rates estimated for the current
dataset, rather than ground-truth channel parameters.

Because insertions and deletions directly induce phase displacement, we further
summarize their combined effect using the phase-load proxy
\begin{equation}
\Lambda=L(\hat{p}_i+\hat{p}_d),
\label{eq:phase_load}
\end{equation}
which characterizes the estimated indel burden over a target-length strand and
is used to adapt the subsequent decoding scope.

The decoding scope is determined jointly from $\rho$, $\Lambda$, and $a_i$;
full routing rules are provided in the Supplementary Methods.

\subsection{Evidence-Guided Phase Dynamic Programming}
\label{sec:phase_dp}

The provisional consensus already provides accurate reconstruction for most
clusters. Under low coverage or elevated indel rates, however, residual
errors or phase ambiguity may remain in a subset of sequences. CCP-DP
therefore selectively refines clusters identified by the adaptive routing
procedure. Sequences exhibiting a clear localized boundary phase shift are
corrected using recurring terminal structures observed across clusters,
whereas more general phase ambiguities are handled by Phase-DP. The two
refinement routes are mutually exclusive, so each routed cluster is processed
by at most one structured correction.

For the boundary route, a candidate correction is accepted only when
$G_{\mathrm{read}}+A_{\mathrm{anchor}}>0$, where $G_{\mathrm{read}}$ is the
reduction in read-edit energy and $A_{\mathrm{anchor}}$ the restored terminal-anchor
support; otherwise, $g_i$ is retained.

For clusters that remain phase-ambiguous, CCP-DP formulates the subsequent
reconstruction as a length-constrained maximum-score path inference problem
over a phase-aware trellis, where sequence variations and phase shifts
induced by IDS errors are jointly represented. For an active cluster, let
$R=|g_i|$ denote the provisional-consensus length and $L$ the target length.
The dynamic-programming state is defined as
\begin{equation}
D(r,\ell,h,z),
\end{equation}
where $r$ and $\ell$ denote the current reference position and generated
sequence length, respectively, $h$ represents the recent sequence context,
and $z$ indicates the current insertion state. Match/substitution, deletion,
and insertion transitions update $(r,\ell)$ as $(r+1,\ell+1)$,
$(r+1,\ell)$, and $(r,\ell+1)$, respectively. To exclude implausible phase
drift, decoding is restricted to a bounded phase band, and valid paths must
terminate at $(R,L)$. The exact phase-band constraint is given in the
Supplementary Methods.

Each transition $e$ is scored by integrating three complementary sources
of evidence:
\begin{equation}
S(e)
=
S_{\mathrm{read}}(e)
+
\lambda_c^{\mathrm{eff}} S_{\mathrm{ctx}}(e)
+
0.10\,S_{\mathrm{IDS}}(e),
\label{eq:phase_score}
\end{equation}
where $S_{\mathrm{read}}$ summarizes the cluster-local evidence obtained from
read--guide alignments together with local and position-aware sequence
support. $S_{\mathrm{ctx}}$ represents the forward--reverse Cross-LOO
contextual support modulated by the remaining clusters, whereas
$S_{\mathrm{IDS}}$ is the log-likelihood of the corresponding
match/substitution, insertion, or deletion event under the estimated blind
IDS profile. The effective cross-cluster weight is defined as
\begin{equation}
\lambda_c^{\mathrm{eff}}=\tau\lambda_c^{0},
\end{equation}
so the contribution of cross-cluster evidence is controlled by its
dataset-level reliability. The dynamic-programming update is
\begin{equation}
D(r',\ell',h',z')
=
\max_{(r,\ell,h,z)\xrightarrow{e}(r',\ell',h',z')}
\left[D(r,\ell,h,z)+S(e)\right].
\end{equation}

The highest-scoring path satisfying the target-length constraint defines the
Phase-DP candidate. The candidate replaces the provisional consensus only if
it achieves a higher objective score; ties retain the provisional consensus. The complete reconstruction procedure is provided in
Supplementary Algorithm~S1.

\section{Experiments and Results}
\subsection{Experimental Setup}
\label{sec:experimental_setup}

\textbf{Problem setup and preprocessing.}
Each experiment takes clustered noisy reads and the target strand length $L$
as input. CCP-DP reconstructs one length-$L$ sequence per cluster using only
the observed reads and statistics estimated from the input data; reference
sequences are used only for evaluation. Reads are normalized to the A/C/G/T
alphabet before reconstruction.

\textbf{Datasets.}
We evaluate CCP-DP on four public DNA storage sequencing datasets:
Erlich et al.\ (152 bp)~\citep{erlich2017dna},
Srinivasavaradhan et al.\ (110 bp)~\citep{srinivasavaradhan2021trellis},
Chandak et al.\ (108 bp)~\citep{chandak2020nanopore}, and
Chandak et al.\ (150 bp)~\citep{chandak2019ldpc}.
They are hereafter denoted as Erlich-152, Srinivas-110,
Chandak-108, and Chandak-150, respectively.
These datasets differ substantially in sequencing coverage and IDS error
characteristics, providing diverse conditions for evaluating reconstruction
performance.
\begin{figure*}[!t]
    \centering
    \vspace{-0.2in}
    \includegraphics[width=0.90\textwidth]
    {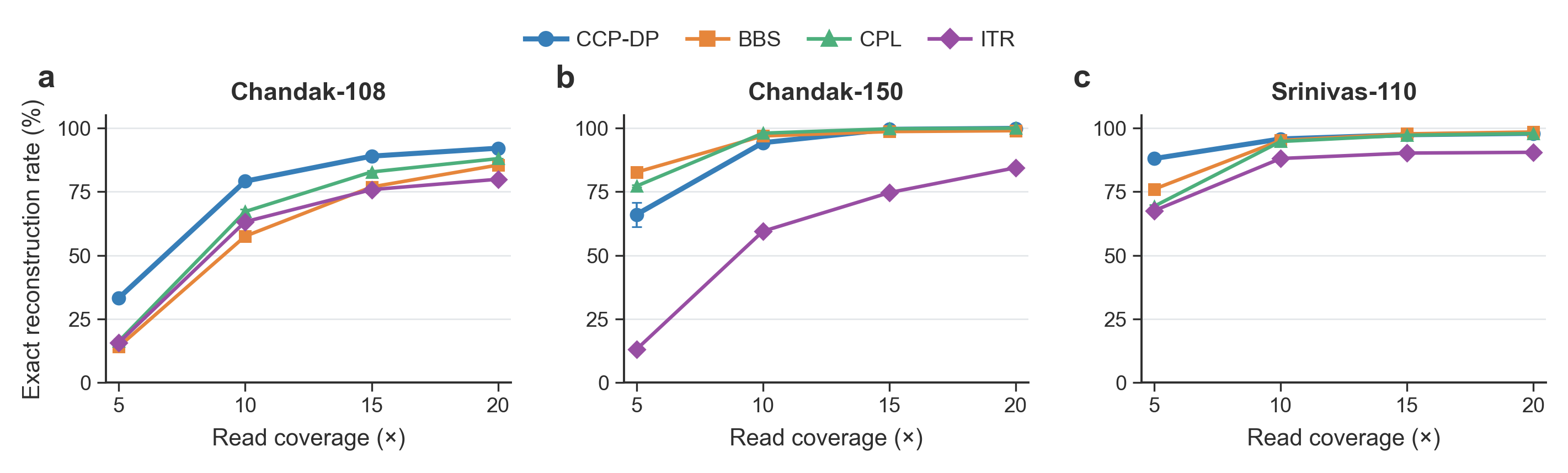}
  \vspace{-0.2in}  \caption{
    \textbf{Exact reconstruction performance across sequencing coverages.}
    Exact reconstruction rates from \(5\times\) to \(20\times\) coverage are
    shown for \textbf{(a)} Chandak-108, \textbf{(b)} Chandak-150, and
    \textbf{(c)} Srinivas-110. CCP-DP is compared with BBS, CPL, and ITR
    under identical sampled read clusters. Points show the mean over five
    random seeds, and error bars indicate standard deviations.
    }
    \label{fig:coverage_robustness}
\end{figure*}

\textbf{Baseline methods.}
We compare CCP-DP with five representative reconstruction baselines:
MUSCLE~\citep{edgar2004muscle},
ITR~\citep{sabary2024reconstruction},
CPL~\citep{barlev2025scalable},
BBS~\citep{gu2025efficient}, and
TrellisBMA~\citep{srinivasavaradhan2021trellis}.
We further compare CCP-DP with the deep-learning-based methods
DNAFormer~\citep{barlev2025scalable} and
TReconLM~\citep{weindel2026trace} on Srinivas-110 and Chandak-108,
for which trained models are available.

\textbf{Evaluation metrics.}
We use the exact reconstruction rate as the primary evaluation metric.
For a test set containing $N$ read clusters, it is defined as follows:
\begin{equation}
\mathrm{Success}
=
\frac{1}{N}
\sum_{i=1}^{N}
\mathbb{I}\!\left(\hat{x}_i=x_i\right)
\times 100\%,
\end{equation}
where $x_i$ and $\hat{x}_i$ denote the original and reconstructed sequences
of the $i$-th cluster, respectively.
Residual reconstruction error is additionally measured by the mean
Levenshtein edit distance (Mean ED), using unit-cost insertion, deletion,
and substitution. Its formal definition is provided in Supplementary Methods.
Runtime is used to assess computational efficiency.

CCP-DP runtime covers the complete reconstruction pipeline. CPU-based methods
use up to 32 threads when supported, whereas DNAFormer and TReconLM runtimes
include inference only; detailed settings are provided in Supplementary
Methods.

\subsection{Overall Reconstruction Performance}
\label{sec:overall_performance}

Table~\ref{tab:main_results} compares CCP-DP with representative
reconstruction baselines on four public datasets. Overall, CCP-DP achieves high reconstruction accuracy across all evaluated datasets. On Erlich-152, where performance is already near saturation, CCP-DP reaches 99.98\% exact reconstruction, tying with BBS for the best
result. On Srinivas-110, Chandak-108, and Chandak-150, CCP-DP achieves the highest exact reconstruction rates of 98.01\%, 95.36\%, and 98.25\%, respectively. The largest improvement over the strongest baseline is observed
on Srinivas-110, where CCP-DP improves exact reconstruction from 95.82\% to
98.01\% while requiring 1.782~s.

\begin{table}[!htbp]
\centering
\caption{Overall reconstruction performance.}
\label{tab:main_results}
\small
\setlength{\tabcolsep}{3pt}
\renewcommand{\arraystretch}{1.05}
\begin{tabular}{llccc}
\toprule
Dataset & Method & Exact Rec. (\%) & Mean ED & Runtime (s) \\
\midrule

\multirow{6}{*}{Erlich-152}
& MUSCLE      & 99.96 & 0.0003 & 197.340 \\
& ITR         & 99.97 & \textbf{0.0002} & 12122.000 \\
& CPL         & 99.97 & 0.0003 & 162.670 \\
& BBS         & \textbf{99.98} & \textbf{0.0002} & 12.950 \\
& TrellisBMA  & 99.96 & 0.0007 & 8827.000 \\
& \textbf{CCP-DP}
              & \textbf{99.98} & \textbf{0.0002} & \textbf{7.456} \\
\midrule

\multirow{6}{*}{Srinivas-110}
& MUSCLE      & 73.81 & 0.3745 & 409.898 \\
& ITR         & 87.58 & 0.2323 & 386.570 \\
& CPL         & 94.85 & 0.1540 & 25.883 \\
& BBS         & 95.82 & 0.1298 & 3.060 \\
& TrellisBMA  & 69.81 & 0.8811 & 1373.700 \\
& \textbf{CCP-DP}
              & \textbf{98.01} & \textbf{0.0491} & \textbf{1.782} \\
\midrule

\multirow{6}{*}{Chandak-108}
& MUSCLE      & 62.76 & 0.5921 & 3185.340 \\
& ITR         & 64.12 & 0.6657 & 87.470 \\
& CPL         & 90.93 & 0.2025 & 162.670 \\
& BBS         & 93.45 & 0.2019 & \textbf{1.110} \\
& TrellisBMA  & 52.11 & 1.7094 & 866.490 \\
& \textbf{CCP-DP}
              & \textbf{95.36} & \textbf{0.1221} & 2.200 \\
\midrule

\multirow{6}{*}{Chandak-150}
& MUSCLE      & 36.44 & 0.7148 & 170.560 \\
& ITR         & 72.19 & 0.3431 & 2002.900 \\
& CPL         & 97.81 & 0.0588 & 30.380 \\
& BBS         & 97.01 & 1.7819 & 3.780 \\
& TrellisBMA  & 66.34 & 0.9508 & 1432.890 \\
& \textbf{CCP-DP}
              & \textbf{98.25} & \textbf{0.0520} & \textbf{1.827} \\
\bottomrule
\end{tabular}
\end{table}

Comparisons with the deep-learning-based methods DNAFormer and TReconLM are
reported in Supplementary Table~S2. CCP-DP achieves higher reconstruction
accuracy on both Srinivas-110 and Chandak-108, without requiring
dataset-specific training or fine-tuning.

\subsection{Robustness Across Sequencing Coverages}
\label{sec:coverage_robustness}

\begin{figure*}[!t]
    \centering

    \includegraphics[width=0.92\textwidth]
    {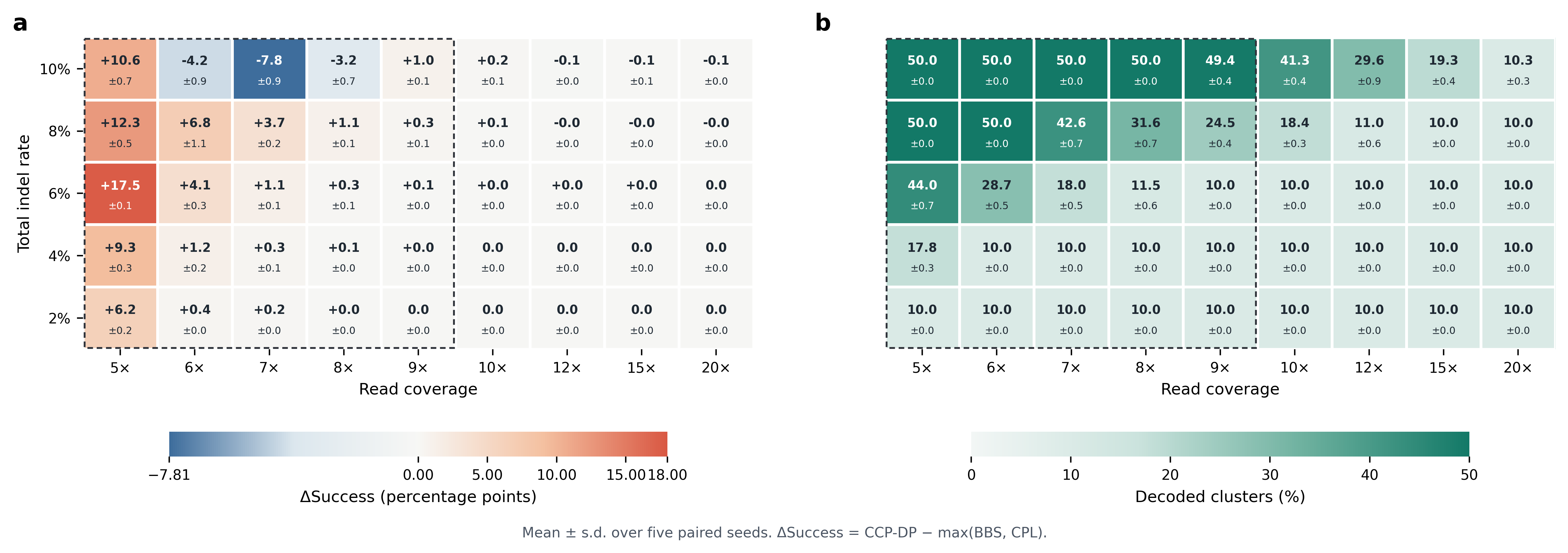}
    \vspace{-0.1in}
    \caption{
\textbf{Performance across coverage and IDS conditions.}
\textbf{(a)} Exact-reconstruction gain over the stronger of BBS and CPL
(mean $\pm$ s.d., five seeds).
\textbf{(b)} Fraction of clusters routed to structured decoding;
dashed boxes mark the densely sampled low-coverage regime.
}
    \label{fig:coverage_ids}

    \vspace{0.12in}

    \includegraphics[width=0.92\textwidth]
    {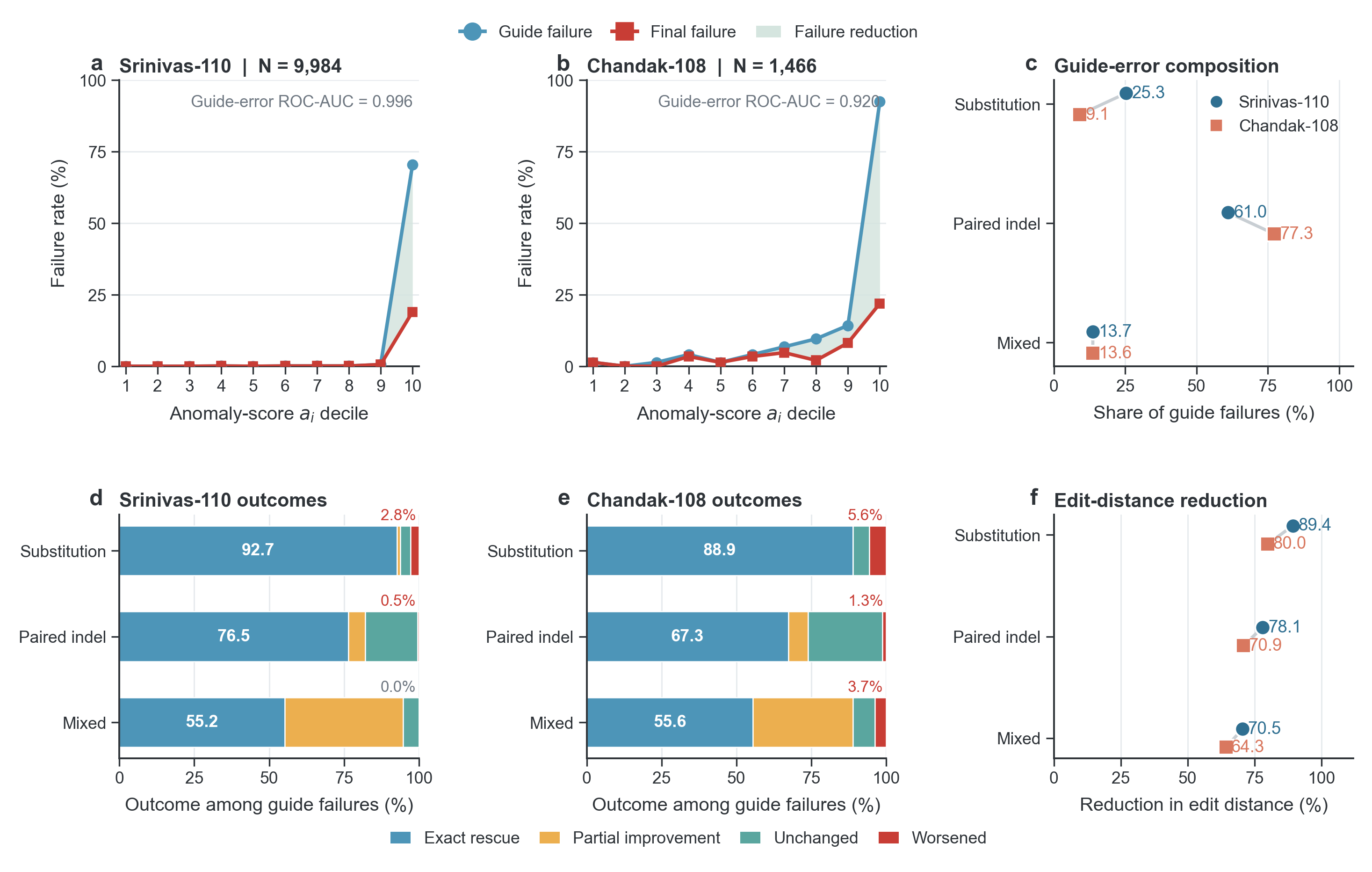}
    \vspace{-0.2in}
    \caption{
    \textbf{Difficult-cluster identification and error correction in CCP-DP.}
    \textbf{(a,b)} Guide and final failure rates across anomaly-score deciles
    for Srinivas-110 and Chandak-108.
    \textbf{(c)} Error-type composition of erroneous provisional guides.
    \textbf{(d,e)} Reconstruction outcomes after structured decoding.
    \textbf{(f)} Relative edit-distance reduction by guide-error type.
    }
    \label{fig:mechanism}

\end{figure*}

To evaluate reconstruction robustness under limited sequencing coverage,
we tested Chandak-108, Chandak-150, and Srinivas-110 from \(5\times\) to
\(20\times\) coverage using identical sampled read clusters across methods.
As shown in Fig.~\ref{fig:coverage_robustness}, the advantage of CCP-DP is most
pronounced in the low-coverage regime. CCP-DP maintains a clear lead
on Chandak-108 and remains competitive on Chandak-150 and Srinivas-110 as
coverage increases. The corresponding mean edit-distance results are provided
in Supplementary Fig.~S4.

\subsection{Performance Across Coverage and IDS Conditions}
\label{sec:coverage_ids}

To further characterize CCP-DP under varying coverage and IDS conditions,
synthetic read clusters were generated from Srinivas-110 across nine
coverage levels ($5$--$20\times$) and five total indel rates (2--10\%),
with a fixed 2\% substitution rate and equal insertion and deletion rates.
Each condition used five paired random seeds, and CCP-DP was compared with
BBS and CPL on identical generated inputs.

As shown in Fig.~\ref{fig:coverage_ids}(a), the advantage of CCP-DP is most
pronounced at low coverage. At $5\times$ coverage, CCP-DP
improves exact reconstruction by 6.2--17.5 percentage points over
the stronger of BBS and CPL, with the largest gain of
$17.5\pm0.1$ percentage points at a 6\% indel rate. As coverage
increases, the local read evidence reduces the
relative gain of CCP-DP. Under the extreme 10\% indel condition, negative
gains appear at several intermediate coverage levels, indicating remaining limitations under severe IDS corruption.

Fig.~\ref{fig:coverage_ids}(b) further shows that the routed fraction
tracks reconstruction difficulty, increasing as coverage decreases and
the indel rate rises.

\subsection{Difficult-Cluster Identification and Error Correction}
\label{sec:mechanism_analysis}

Because most clusters are already resolved during provisional
consensus construction, the remaining errors are concentrated in a smaller subset
of difficult clusters. As shown in Fig.~\ref{fig:mechanism}(a,b), guide
failures are strongly enriched in the highest anomaly-score deciles, with
ROC-AUC values of 0.996 and 0.920 on Srinivas-110 and Chandak-108,
respectively. Structured decoding substantially reduces the corresponding
failure rates, supporting anomaly-guided selective refinement.

Error-type analysis shows that paired indels dominate the erroneous
provisional guides, accounting for 61.0\% and 77.3\% of guide failures on
Srinivas-110 and Chandak-108, respectively
(Fig.~\ref{fig:mechanism}(c)). Structured decoding exactly rescues 76.5\%
and 67.3\% of these paired-indel errors and also reduces edit distance across
substitution, paired-indel, and mixed-error categories
(Fig.~\ref{fig:mechanism}(d--f)). These results show that CCP-DP concentrates refinement on difficult
clusters and effectively corrects the dominant residual error modes.

\subsection{Cross-Cluster Context and Selective Decoding}
\label{sec:cross_context_adaptive}

\begin{figure*}[!t]
    \centering
    \includegraphics[width=0.84\textwidth]
    {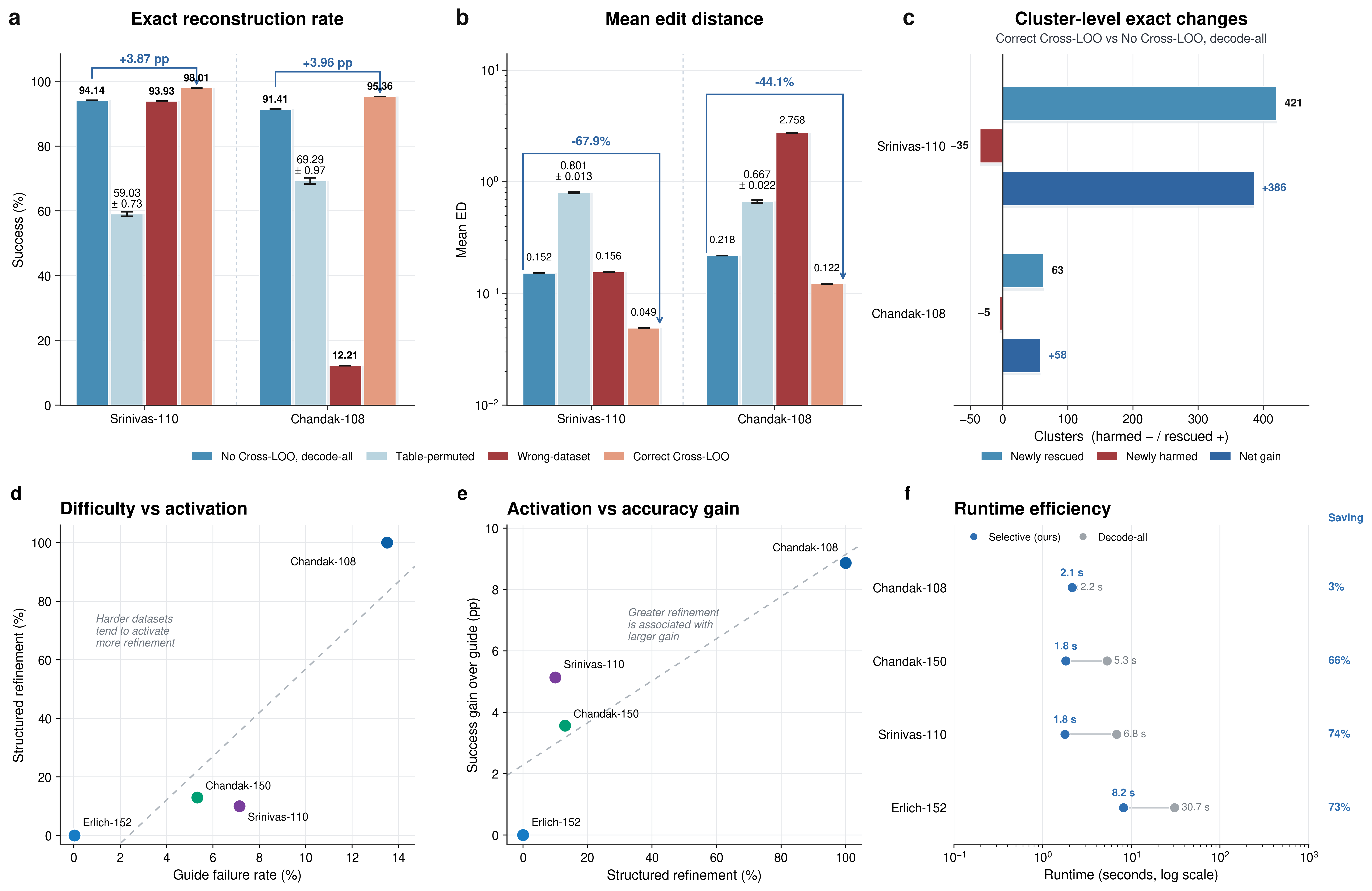}
  \vspace{-0.1in}  \caption{
\textbf{Cross-cluster context and selective decoding in CCP-DP.}
\textbf{(a,b)} Exact reconstruction and mean edit distance under four
context settings on Srinivas-110 and Chandak-108.
\textbf{(c)} Cluster-level changes induced by Correct Cross-LOO relative
to No Cross-LOO, decode-all.
\textbf{(d,e)} Relationships between decoding fraction, provisional-guide
failure, and reconstruction gain.
\textbf{(f)} Runtime comparison between selective and decode-all decoding.
}
    \label{fig:cross_context_adaptive}
\end{figure*}

To isolate the contribution of cross-cluster context, we compared Correct Cross-LOO with the No Cross-LOO, decode-all, Table-permuted, and Wrong-dataset controls on
Srinivas-110 and Chandak-108. All four conditions decoded every cluster using the same guides, blind IDS profile, calibration factor and decoder settings; they differed only in the Cross-LOO context supplied to the trellis. Detailed definitions are provided in the Supplementary Methods.

As shown in Fig.~\ref{fig:cross_context_adaptive}(a--c), Correct Cross-LOO
improves exact reconstruction relative to
No Cross-LOO, decode-all, from 94.14\% to 98.01\% on
Srinivas-110 and from 91.41\% to 95.36\% on Chandak-108, while also
substantially reducing the mean edit distance. In contrast, permuted or
mismatched contextual information consistently degrades reconstruction
performance. At the cluster level, Correct Cross-LOO yields net gains of
386 and 58 exact reconstructions on the two datasets, respectively. Together,
these
results show that the benefit of Cross-LOO arises from reusable
within-dataset contextual structure rather than from merely introducing
an additional scoring term.

The value of selective decoding is further reflected in how CCP-DP
adapts its decoding scope across datasets. Rather than fixing the amount
of structured decoding in advance, CCP-DP adjusts the routed fraction
according to provisional-guide reliability and dataset-level error
conditions. As shown in Fig.~\ref{fig:cross_context_adaptive}(d--e),
Erlich-152 requires almost no further refinement, whereas approximately
10\% and 13\% of clusters are routed on Srinivas-110 and Chandak-150,
respectively. In contrast, the more challenging Chandak-108 dataset
activates structured decoding for all clusters.

This adaptive allocation also improves efficiency. Relative to
decode-all reconstruction, selective routing reduces runtime by
approximately 66--74\% on Erlich-152, Srinivas-110, and Chandak-150,
whereas little additional saving is possible on Chandak-108 because
nearly all clusters require further decoding
(Fig.~\ref{fig:cross_context_adaptive}(f)).

\subsection{Ablation Study}
\label{sec:ablation}

Ablation experiments further verify the roles of the main CCP-DP components.
On Srinivas-110, removing the Cross-LOO score while retaining adaptive
routing reduced exact reconstruction from 98.01\% to 94.14\%; the Local
trellis, decode-all variant, which removes both Cross-LOO and blind IDS
likelihood terms, reached 94.86\%. Removing the blind IDS likelihood yielded
97.94\%. Full decode-all retained 98.01\% accuracy but increased runtime
from 1.782~s to 6.824~s, supporting adaptive selective routing.
Complete ablation and sensitivity results are provided in Supplementary
Table~S3 and Fig.~S3.

\section{Conclusion}

This work shows that reusable sequence context across read clusters can
provide effective complementary evidence for DNA trace reconstruction when
cluster-local observations are insufficient. CCP-DP integrates cluster-local
reconstruction with calibrated cross-cluster context and selective phase-aware
refinement, allowing additional evidence to be introduced when local
reconstruction remains ambiguous. Across multiple real DNA storage datasets
and diverse sequencing conditions, the method achieves consistently high exact
reconstruction accuracy while maintaining low residual error and computational
cost. Mechanism and ablation analyses further demonstrate that the observed
gains arise primarily from informative cross-cluster context, while selective
decoding reduces computation by concentrating refinement on clusters with
greater residual uncertainty.

The benefit of CCP-DP depends on the strength of reusable cross-cluster
structure and the fraction of clusters requiring structured decoding.
When cross-cluster patterns are weak or difficult clusters are widespread,
its statistical and computational advantages may be reduced. Future work
may explore more flexible cross-cluster modeling, improved confidence
estimation, and broader adaptation to diverse sequencing conditions.

\section*{Data and Code Availability}

The datasets analyzed in this study are publicly available from the
corresponding original publications. The CCP-DP implementation and scripts
for reproducing the main experiments are available at
\url{https://github.com/ZixinZ3899/CCP-DP}; no new sequencing data were
generated in this study.

\section*{Funding}
This work is supported by the National Key Research and Development Program of China (2025YFC3409900), the Tianjin Municipal Science and Technology Major Program (25ZXZSSS00390), the Tianjin Future Industry Science and Technology Major Project (25ZXWCSY00290), the Emerging Frontiers Cultivation Program of Tianjin University Interdisciplinary Center, and the Independent Research Project of State Key Laboratory of Synthetic Biology (HCZC-202605A).


\bibliographystyle{oup-abbrvnat}
\bibliography{references}

\clearpage
\begin{center}
{\Large\bfseries Supplementary Materials for\par}
\vspace{6pt}
{\LARGE\bfseries CCP-DP: Cross-cluster context-guided phase dynamic programming for DNA storage sequence reconstruction\par}
\vspace{8pt}
{\normalsize Zixin Zhang$^{1}$, Guanjin Qu$^{1}$, and Huaming Wu$^{1,2,*}$\par}
\vspace{5pt}
{\small $^{1}$Center for Applied Mathematics, KL-AAGDM, Tianjin University, China\par
$^{2}$State Key Laboratory of Synthetic Biology, Tianjin University, China\par
$^{*}$Corresponding author: \href{mailto:whming@tju.edu.cn}{whming@tju.edu.cn}\par}
\end{center}
\vspace{12pt}
\setcounter{section}{0}
\setcounter{subsection}{0}
\setcounter{figure}{0}
\setcounter{table}{0}
\setcounter{equation}{0}
\setcounter{algorithm}{0}
\renewcommand{\thesection}{S\arabic{section}}
\renewcommand{\thesubsection}{\thesection.\arabic{subsection}}
\renewcommand{\thefigure}{S\arabic{figure}}
\renewcommand{\thetable}{S\arabic{table}}
\renewcommand{\theequation}{S\arabic{equation}}
\renewcommand{\thealgorithm}{S\arabic{algorithm}}
\renewcommand{\figurename}{Supplementary Figure}
\renewcommand{\tablename}{Supplementary Table}
\section{Supplementary Methods}

\subsection{Implementation details of provisional consensus construction}
\label{sec:supp_provisional}

\paragraph{High-coverage positional $k$-mer phase chain.}

For high-coverage clusters, the abundance of read observations makes
position-specific $k$-mer statistics sufficiently stable to construct the
provisional guide without requiring a full multiple-read alignment.
The high-coverage branch therefore represents each read as a sequence of
overlapping $k$-mers and maps these local sequence patterns onto a common
target coordinate system.

Let $r_j$ be a read of length $n_j$, and let the $i$-th $k$-mer in $r_j$
start at read coordinate $i$. Its center target layer is defined as
\begin{equation}
p_j(i)
=
\operatorname{round}
\left(
\frac{i(L-k)}{n_j-k}
\right),
\label{eq:supp_kmer_projection}
\end{equation}
where $L$ is the target sequence length. This normalization maps the
$k$-mer layers of reads with different lengths onto the common target
layer range $0,\ldots,L-k$.

Because insertions and deletions can introduce small local coordinate
shifts, a projected $k$-mer does not contribute evidence only to its
center layer. Instead, its support is distributed over a local
neighborhood around $p_j(i)$. With the default positional radius
$r=2$, the support contribution at offset $\Delta$ is
\begin{equation}
w(\Delta)
=
r+1-|\Delta|,
\qquad |\Delta|\le r,
\label{eq:supp_kmer_weight}
\end{equation}
which correspond to relative weights $1,2,3,2,1$ over layers
$p-2,p-1,p,p+1,p+2$, respectively. Positions outside the valid target
layer range are ignored. This local spreading allows evidence from the
same underlying sequence context to remain mutually supportive even when
small indel-induced shifts cause corresponding $k$-mers from different
reads to be projected to slightly different target layers.

The neighborhood-weighted support is then accumulated across all reads in
the cluster. At each target layer, multiple $k$-mer candidates may therefore
receive different levels of aggregated support. Rather than selecting the
highest-supported candidate independently at each position, CCP-DP links
candidates at adjacent layers only when the suffix of length $k-1$ of the
preceding $k$-mer matches the prefix of length $k-1$ of the succeeding
$k$-mer. These overlap constraints form a layered $k$-mer phase chain that
preserves sequence continuity across the target.

Dynamic programming is finally used to identify the overlap-consistent path
with the maximum cumulative support across the $L-k+1$ target layers.
The first $k$-mer initializes the reconstructed sequence, and each subsequent
layer contributes the final nucleotide of the selected $k$-mer, yielding a
provisional guide of exactly length $L$.

\begin{figure}[t]
    \centering
    \includegraphics[width=0.90\textwidth]
    {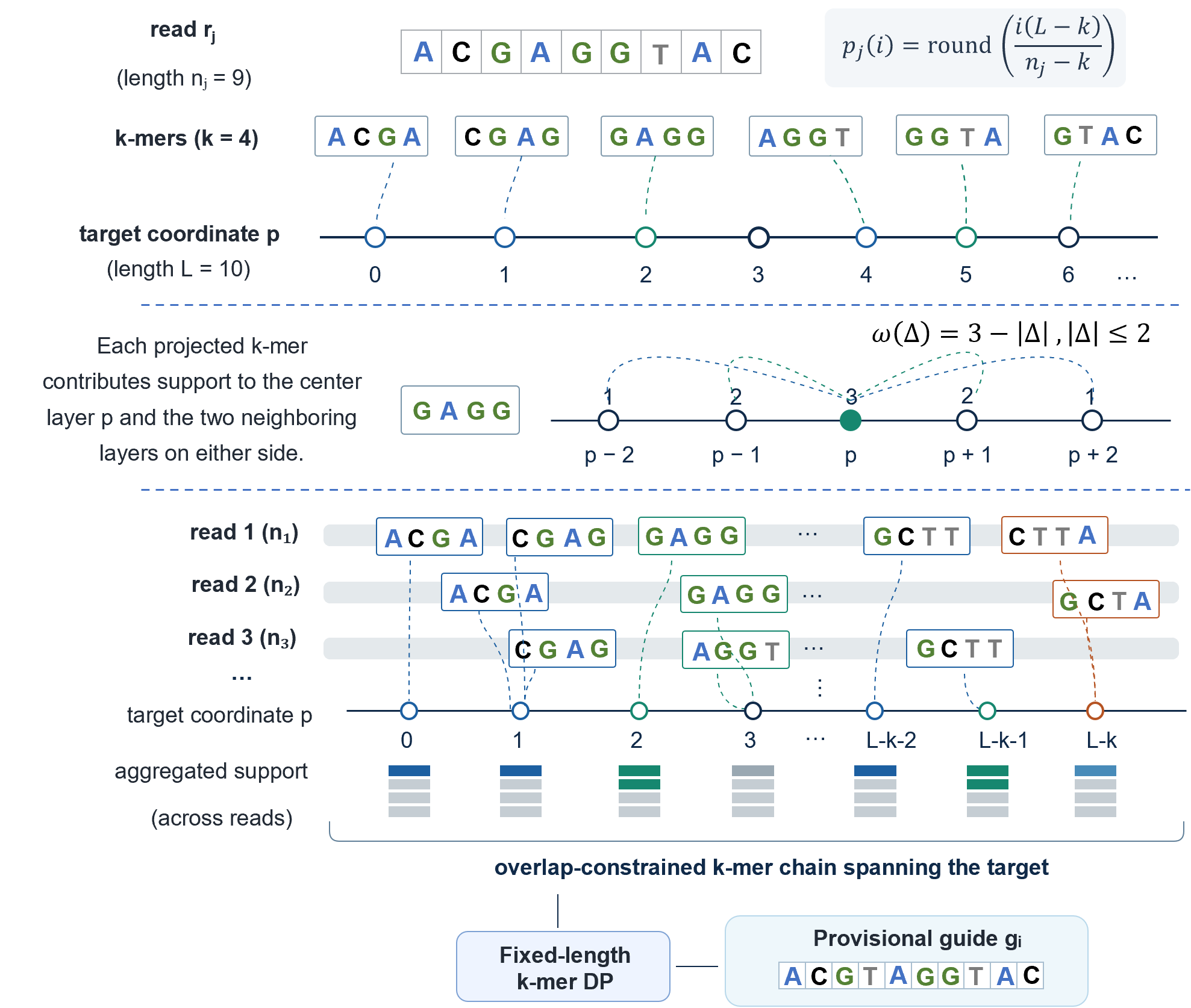}
    \caption{
    \textbf{Position-aware $k$-mer aggregation and fixed-length decoding
    in the high-coverage provisional-consensus branch.}
    Overlapping $k$-mers extracted from each read are first mapped to
    normalized target layers. Each projected $k$-mer contributes
    distance-weighted support to a local positional neighborhood, and
    support is accumulated across reads. Candidate $k$-mers at adjacent
    layers are connected through $(k-1)$-base overlap constraints, and
    dynamic programming recovers the maximum-support target-length path.
    A schematic example with $n_j=9$, $L=10$, and $k=4$ is shown for
    illustration.
    }
    \label{fig:supp_high_coverage}
\end{figure}

Supplementary Fig.~\ref{fig:supp_high_coverage} illustrates this procedure
using a schematic example with $n_j=9$, $L=10$, and $k=4$. The example
values are chosen only for visual clarity.

\paragraph{Low-coverage banded-evidence reconstruction.}

For lower-coverage clusters, the representative read is used mainly as a
local coordinate reference. It is not treated as a fixed consensus.
The selected reads are aligned independently to this reference using
banded global edit-distance alignment. After alignment, equivalent gap
positions in repetitive regions are left-normalized. This prevents
evidence for the same indel event from being split across neighboring
positions.

The alignments are then converted into two types of local evidence.
At each seed position $j$, $B_j$ records the support for $A$, $C$, $G$,
$T$, and deletion. Between adjacent seed positions, $I_j$ records
insertion candidates, including the empty event for no insertion.
For a seed of length $m$, these evidence units are arranged as
\begin{equation}
I_0,B_0,I_1,B_1,\ldots,B_{m-1},I_m ,
\label{eq:supp_low_coverage_layers}
\end{equation}
which gives $2m+1$ ordered evidence layers. The aggregated counts are converted
into smoothed log-scores before decoding. Mild priors favor no insertion
and no deletion, while gap events receive an additional penalty.

The evidence layers are decoded jointly under the known target length $L$.
Let $D_s(\ell)$ denote the best cumulative score after the first $s$ layers
have been processed and $\ell$ nucleotides have been generated. Emitting a
base from $B_j$ increases the output length by one, whereas deletion leaves
it unchanged. Likewise, an insertion from $I_j$ increases the output length,
while no insertion keeps the same length. States exceeding $L$ are
discarded. The provisional guide is recovered by backtracking from the best
terminal state with length exactly $L$. Thus, the method does not make
independent decisions at each position. Instead, it searches for the
highest-scoring sequence of local events that is globally consistent with
the target length.

\begin{figure}[ht!]
    \centering
    \includegraphics[width=0.98\textwidth]
    {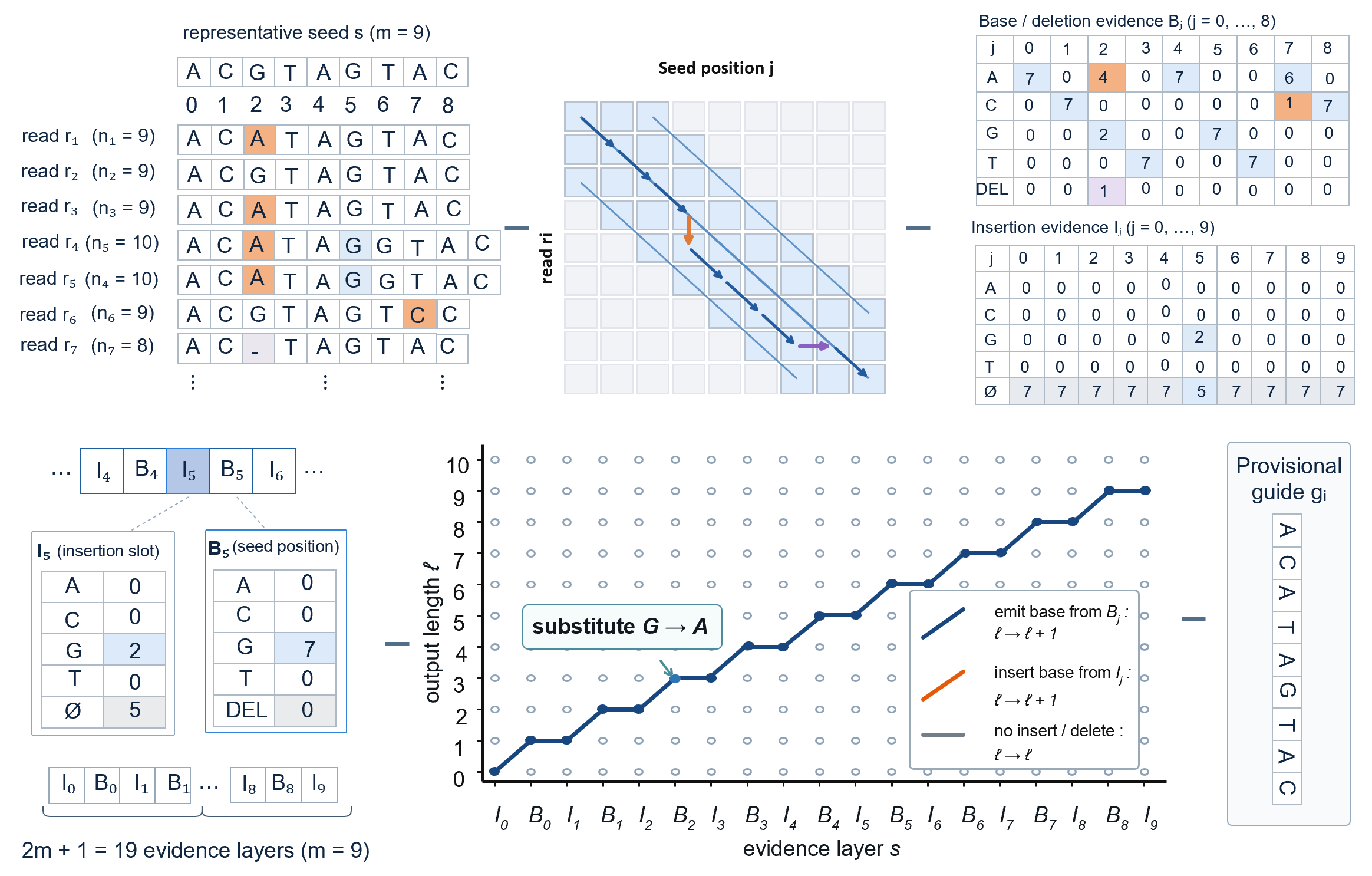}
    \caption{
    \textbf{Low-coverage provisional-consensus construction using
    alignment-derived evidence and fixed-length dynamic programming.}
    Selected reads are aligned to a representative seed, and the resulting
    alignments are summarized into base/deletion evidence $B_j$ at seed
    positions and insertion evidence $I_j$ between adjacent positions.
    These evidence units are arranged as
    $I_0,B_0,\ldots,B_{m-1},I_m$ and jointly decoded under the target-length
    constraint. In the schematic example ($m=L=9$), evidence at $B_2$
    corrects the representative-seed nucleotide from $G$ to $A$, whereas
    the insertion candidate $G$ at $I_5$ is not selected by the optimal
    length-9 path. Backtracking yields the provisional guide
    $\mathrm{ACATAGTAC}$.
    }
    \label{fig:supp_low_coverage}
\end{figure}

Supplementary Fig.~\ref{fig:supp_low_coverage} illustrates the process with
$m=L=9$. At $B_2$, the representative seed contains $G$, but the aligned
reads provide stronger support for $A$. The DP therefore chooses $A$ and
corrects the seed at this position. At $I_5$, two reads support an inserted
$G$, whereas five reads support no insertion. The insertion remains a valid
candidate, but the best length-9 path does not select it. The final
provisional guide is therefore $\mathrm{ACATAGTAC}$. This example shows
that the representative read provides the coordinate system, while the combined read evidence and the
target-length constraint determine the
final sequence.

\subsection{Cross-LOO trust calibration}
\label{sec:supp_crossloo_calibration}

Because the reliability of cross-cluster context can vary across datasets,
CCP-DP uses a dataset-level trust coefficient $\tau \in [0,1]$ to adaptively
control the contribution of Cross-LOO evidence. The coefficient determines
the degree to which the Cross-LOO distribution is shrunk toward the uniform
distribution:
\begin{equation}
\widetilde{P}_{-i}(b\mid u;\tau)
=
(1-\tau)\frac{1}{4}
+
\tau P_{-i}(b\mid u).
\label{eq:supp_crossloo_shrinkage}
\end{equation}
Smaller values of $\tau$ reduce the influence of cross-cluster context,
whereas larger values preserve more of the dataset-specific Cross-LOO
information.

To select this coefficient, CCP-DP evaluates candidate values from
$\{0,0.125,\ldots,1\}$ using separate calibration and validation subsets
formed from provisional-guide transitions. The candidate with the largest
mean predictive gain on the calibration subset is selected provisionally,
and its stability is then assessed on the validation subset. It is retained
only when the validation gain exceeds
\begin{equation}
\max(0.005,\,2\,\mathrm{SE}),
\label{eq:supp_crossloo_acceptance}
\end{equation}
where $\mathrm{SE}$ denotes the standard error of the validation gain.
Otherwise, $\tau$ is set to zero, thereby suppressing unreliable
cross-cluster evidence. The entire calibration procedure relies only on
provisional guides and statistics estimated from the current input, without
using ground-truth sequences.

\subsection{Adaptive routing of structured decoding}
\label{sec:supp_routing}

CCP-DP does not apply structured decoding uniformly to all read clusters.
Instead, the decoding scope is determined adaptively from dataset-level
evidence and cluster-specific uncertainty. The routing policy is controlled
primarily by the cross-cluster reuse rate $\rho$, the phase-load proxy
$\Lambda$, and the cluster anomaly score $a_i$. All of these quantities are
computed from provisional consensuses, observed reads, and dataset-level
statistics, without access to the ground-truth sequences.

\paragraph{Cluster anomaly score.}

For each provisional consensus $g_i$, CCP-DP measures local inconsistency
using both cross-cluster contextual evidence and within-cluster read evidence.
For position $j$, let $c_{i,j}$ denote the Cross-LOO context anomaly and
$\ell_{i,j}$ the within-cluster read-context anomaly. These two components
are combined as
\begin{equation}
q_{i,j}
=
c_{i,j}
+
\lambda_{\mathrm{loc}}\ell_{i,j},
\label{eq:supp_position_anomaly}
\end{equation}
where $\lambda_{\mathrm{loc}}$ controls the contribution of the
within-cluster component. Both anomaly terms are derived from the maximum
surprise score among the context windows covering position $j$, so that
pronounced local inconsistencies are not diluted by neighboring
well-supported regions.

To summarize the uncertainty of the entire cluster, CCP-DP averages the
$K_i$ most anomalous positions:
\begin{equation}
a_i
=
\frac{1}{K_i}
\sum_{j\in\operatorname{TopK}(q_i)}
q_{i,j},
\qquad
K_i=\min(24,L).
\label{eq:supp_cluster_anomaly}
\end{equation}
A larger $a_i$ therefore indicates a greater concentration of unresolved
local inconsistency in the provisional consensus.

When cross-cluster reuse is sufficiently high ($\rho \geq 0.5$), anomaly
scores are used to identify a high-risk subset of clusters. A two-means
partition is applied to the anomaly-score distribution, and the high-anomaly
component is selected when the two components are sufficiently separated.
The resulting decoding fraction is constrained to the range 10--50\%.
If no clearly separated high-anomaly component is detected, CCP-DP falls
back to decoding only the top 10\% of clusters ranked by anomaly score.

When cross-cluster reuse is lower ($\rho < 0.5$), the decoding scope depends
more strongly on the estimated phase load. For low phase load
($\Lambda < 3$), structured decoding is restricted to clusters with
insufficient local support. Under moderate phase load, anomaly and coverage
information are used jointly to broaden the selected subset. When the phase
load becomes sufficiently large ($\Lambda \geq 6$), structured decoding is
applied densely to account for widespread phase ambiguity. In the uncommon
case where the Cross-LOO prior is rejected under low reuse and moderate phase
load, decoding is conservatively expanded to all clusters.

This adaptive allocation concentrates additional computation on difficult
clusters while avoiding unnecessary structured inference for clusters whose
provisional consensuses are already well supported by local evidence.

\subsection{Complete Evidence-Guided Reconstruction Procedure}
\label{sec:supp_phase_dp_algorithm}

The complete reconstruction procedure of CCP-DP is summarized in
Supplementary Algorithm~\ref{alg:supp_phase_dp}. For each cluster, the
algorithm first determines whether a localized boundary correction is
applicable. Otherwise, the routing decision determines whether the
provisional consensus is retained directly or further refined by Phase-DP.
The Phase-DP candidate is accepted only when it achieves a higher score than
the provisional consensus under the same objective.

For an active cluster, let $R=|g_i|$ denote the provisional-consensus
length and $L$ the target length. To restrict implausible phase drift during
Phase-DP, only states satisfying
\begin{equation}
\left|
\ell-\operatorname{round}\left(\frac{rL}{R}\right)
\right|
\leq \beta
\label{eq:supp_phase_band}
\end{equation}
are retained, where $\beta$ denotes the phase-band width. Valid
reconstruction paths must terminate at $(R,L)$, ensuring that the decoded
sequence satisfies the target-length constraint.

Each admissible transition $e$ is scored by combining cluster-local,
cross-cluster, and channel evidence:
\begin{equation}
S(e)
=
S_{\mathrm{read}}(e)
+
\lambda_c^{\mathrm{eff}}S_{\mathrm{ctx}}(e)
+
0.10\,S_{\mathrm{IDS}}(e),
\label{eq:supp_phase_score}
\end{equation}
where $S_{\mathrm{read}}(e)$ denotes the cluster-local evidence derived
from read--guide alignments and position-aware sequence support,
$S_{\mathrm{ctx}}(e)$ denotes the forward--reverse Cross-LOO contextual
support, and $S_{\mathrm{IDS}}(e)$ is the log-likelihood of the
corresponding transition under the estimated blind IDS profile.
The effective cross-cluster weight is
\begin{equation}
\lambda_c^{\mathrm{eff}}
=
\tau\lambda_c^{0},
\label{eq:supp_effective_cross_weight}
\end{equation}
so that the Cross-LOO contribution is modulated by the dataset-level
trust coefficient $\tau$.

\begin{algorithm}[t]
\caption{Evidence-Guided Phase Reconstruction}
\label{alg:supp_phase_dp}
\normalsize
\begin{algorithmic}[1]
\Require Read cluster $R_i$, provisional consensus $g_i$, target length $L$;
         Cross-LOO model $\mathcal{C}_{-i}$; blind IDS profile
         $\hat{\boldsymbol{\theta}}_{\rm IDS}$; routing decision $d_i$
\Ensure Reconstructed sequence $\hat{x}_i$

\If{a localized boundary phase shift is detected}
    \State Construct the boundary-correction candidate $\tilde{x}_i$
    \State Evaluate $G_{\rm read}$ and $A_{\rm anchor}$
    \If{$G_{\rm read}+A_{\rm anchor}>0$}
        \State \Return $\tilde{x}_i$
    \Else
        \State \Return $g_i$
    \EndIf
\EndIf

\If{$d_i=0$}
    \State \Return $g_i$
\EndIf

\State Initialize $D(0,0,h_{\emptyset},0)\gets0$

\For{each valid state $(r,\ell,h,z)$ within the phase band}
    \For{each admissible transition
          $e\in\{\mathrm{M/S},\mathrm{I},\mathrm{D}\}$}
        \State Determine the successor state $(r',\ell',h',z')$
        \State Compute $S(e)$ using the Phase-DP scoring function
        \State Update the corresponding DP state
    \EndFor
\EndFor

\State Recover the Phase-DP candidate $\hat{x}_i$
\State Evaluate $g_i$ under the same scoring objective

\If{$S(\hat{x}_i)>S(g_i)$}
    \State \Return $\hat{x}_i$
\Else
    \State \Return $g_i$
\EndIf

\end{algorithmic}
\end{algorithm}

\subsection{Cross-LOO counterfactual controls}
\label{sec:supp_crossloo_controls}

To isolate the contribution of cross-cluster contextual information, we
performed a counterfactual evaluation on Srinivas-110 and Chandak-108 using
four context configurations: No Cross-LOO, decode-all, Table-permuted, Wrong-dataset, and
Correct Cross-LOO. All configurations used the same target read clusters,
provisional guides, blind IDS profile, decoder parameters, and target length.
Adaptive routing was disabled using the decode-all setting, so every cluster was subjected to structured decoding in each condition.
Ground-truth centers were used only for post hoc evaluation.

The four configurations differed only in the contextual evidence supplied to
the decoder. Correct Cross-LOO used the standard leave-one-cluster-out model
constructed from the provisional guides of the target dataset. No Cross-LOO,
decode-all
disabled only the Cross-LOO contribution to trellis scoring while
retaining the local read evidence, blind IDS likelihood, and all other decoder
components. Table-permuted randomly reassigned the empirical Cross-LOO
statistics across context keys, preserving the overall count distribution
while disrupting the correspondence between sequence context and conditional
statistics. Wrong-dataset instead supplied a Cross-LOO model constructed
from the other real dataset.

For all counterfactual controls, $\tau$ was fixed to the calibrated value
obtained from the correctly matched target-dataset Cross-LOO model.
Thus, the Table-permuted and Wrong-dataset conditions changed only the
Cross-LOO context itself, while keeping its calibrated contribution unchanged. The Table-permuted condition was repeated with five
fixed random seeds (2026--2030), whereas the other three configurations were
deterministic.

\begin{table}[t]
\centering
\caption{
Definition of the Cross-LOO counterfactual controls.
All four conditions decode every cluster and differ only in the
Cross-LOO context supplied to the structured decoder.
}
\label{tab:supp_crossloo_controls}

\small
\setlength{\tabcolsep}{3pt}
\renewcommand{\arraystretch}{1.12}
\renewcommand{\tabularxcolumn}[1]{m{#1}}

\begin{tabularx}{\textwidth}{@{}>{\hsize=1.1\hsize\raggedright\arraybackslash}X*{4}{>{\hsize=0.7\hsize\centering\arraybackslash}X}>{\hsize=2.1\hsize\raggedright\arraybackslash}X@{}}
\toprule
\textbf{Condition} &
\textbf{Guides} &
\textbf{Blind IDS} &
\boldmath$\tau$\unboldmath &
\textbf{Decode set} &
\textbf{Context field} \\
\midrule

No Cross-LOO, decode-all
& Fixed & Fixed & Fixed
& All clusters
& Disabled \\

Table-permuted
& Fixed & Fixed & Fixed
& All clusters
& Permuted target context \\

Wrong-dataset
& Fixed & Fixed & Fixed
& All clusters
& Other-dataset context \\

Correct Cross-LOO
& Fixed & Fixed & Fixed
& All clusters
& Matched target context \\

\bottomrule
\end{tabularx}
\end{table}

\subsection{Experimental protocols and computational settings}
\label{sec:supp_experimental_protocol}

\paragraph{Evaluation metrics.}
In addition to exact reconstruction rate, residual reconstruction error was
measured using the mean Levenshtein edit distance. For $N$ read clusters,
it is defined as
\begin{equation}
\mathrm{Mean\ ED}
=
\frac{1}{N}
\sum_{i=1}^{N}
d_{\mathrm{Lev}}(\hat{x}_i,x_i),
\end{equation}
where $x_i$ and $\hat{x}_i$ denote the original and reconstructed sequences
of the $i$-th cluster, respectively, and $d_{\mathrm{Lev}}$ denotes the
Levenshtein distance with unit-cost insertion, deletion, and substitution.

\paragraph{Input and preprocessing.}
CCP-DP takes as input clustered noisy reads together with the
target strand length $L$. Reconstruction uses only the observed reads,
provisional guides, and dataset-level statistics estimated from the current
input; reference center sequences are used only for post hoc evaluation.
The experiments use processed clustered-read benchmark files. CCP-DP does
not perform additional primer/index trimming or read-orientation
normalization during reconstruction.

\paragraph{Coverage-based experiments.}
For coverage-robustness experiments, the same nested sampled read clusters
were shared by all compared methods, using five random seeds (2026--2030).
CCP-DP was rerun independently on each sampled input so that its provisional
guides, Cross-LOO statistics, blind IDS profile, and routing decisions were
re-estimated from that input. In the controlled coverage--IDS experiments,
synthetic reads were generated from Srinivas-110 under the specified
coverage and IDS settings, with a fixed 2\% substitution rate and equal
insertion and deletion probabilities. All methods were evaluated on the same
generated inputs for each seed and condition.

\paragraph{Computational environment and runtime.}
CPU-based experiments were conducted on a server with two Intel Xeon Gold
6130H processors and 64~GB RAM. CCP-DP was compiled with GCC/G++ 9.4.0
using C++17, \texttt{-O3}, and OpenMP, and was executed with 32 CPU threads.
Deep-learning-based baselines were evaluated on a single NVIDIA RTX 3090 GPU.
CCP-DP runtime is reported as end-to-end program time, including guide
construction, dataset-level statistics, routing, and structured decoding,
but excluding post hoc evaluation. Runtime repetition protocols differed
across experiments: routing and ablation comparisons used three repeated
runs, whereas coverage-based experiments were summarized over five random
seeds.

%


\section{Supplementary Results}

\subsection{Comparison with deep-learning-based baselines}
\label{sec:supp_learning_baselines}

We further compared CCP-DP with the deep-learning-based reconstruction methods
DNAFormer and TReconLM on Srinivas-110 and Chandak-108. The comparison was
restricted to these two datasets because reproducible trained models with
matching strand lengths were available for both deep-learning-based baselines.
Extending the evaluation to datasets with other strand lengths would require
additional model training or fine-tuning, which would introduce
training-dependent variability and complicate direct comparison across
datasets.

In contrast, CCP-DP does not rely on dataset-specific model training.
Its reconstruction procedure adapts to a new input dataset through statistics
estimated directly from the observed read clusters, allowing the same
inference framework to be applied across different target lengths and error
regimes. This provides training-free portability that is not contingent on
the availability of a length-matched trained model.

As shown in Supplementary Table~\ref{tab:supp_learning_baselines}, CCP-DP
achieved the highest exact reconstruction rate and the lowest mean edit
distance on both evaluated datasets. On Srinivas-110, the exact
reconstruction rates of DNAFormer, TReconLM, and CCP-DP were 90.68\%,
92.07\%, and 98.01\%, respectively. On Chandak-108, the corresponding rates
were 34.56\%, 48.19\%, and 95.36\%, demonstrating a substantially larger
performance gap under the more challenging reconstruction condition.

Runtime comparisons should be interpreted with respect to their different
measurement scopes. CCP-DP is reported using end-to-end CPU runtime, whereas
the reported runtimes for the deep-learning-based methods correspond to GPU
inference only and exclude model training and checkpoint loading. Accordingly,
the runtime results are intended to characterize practical computational cost
rather than to imply strictly identical timing protocols across method
classes.

\begin{table}[t]
\centering
\caption{
\textbf{Comparison with deep-learning-based reconstruction methods.}
The comparison is reported on datasets for which reproducible trained models
with matching strand lengths were available for both DNAFormer and TReconLM.
CCP-DP runtime corresponds to end-to-end CPU execution, whereas runtimes of
the deep-learning-based methods correspond to GPU inference only.
}
\label{tab:supp_learning_baselines}

\small
\setlength{\tabcolsep}{3pt}
\renewcommand{\arraystretch}{1.12}
\renewcommand{\tabularxcolumn}[1]{m{#1}}

\begin{tabularx}{\textwidth}{@{}X X c c c c@{}}
\toprule
\textbf{Dataset} &
\textbf{Method} &
\shortstack{\textbf{Exact reconstruction} \\ \textbf{(\%)}} &
\shortstack{\textbf{Reconstruction rate} \\ \textbf{(\%)}} &
\shortstack{\textbf{Mean} \\ \textbf{ED}} &
\shortstack{\textbf{Runtime} \\ \textbf{(s)}} \\
\midrule

\multirow{3}{*}{Srinivas-110}
& DNAFormer
& 90.68
& 99.83
& 0.3277
& 27.32 \\

& TReconLM
& 92.07
& 99.85
& 0.1679
& 330.89 \\

& \textbf{CCP-DP}
& \textbf{98.01}
& \textbf{99.96}
& \textbf{0.0491}
& 1.782 \\
\midrule

\multirow{3}{*}{Chandak-108}
& DNAFormer
& 34.56
& 93.68
& 2.6741
& 8.97 \\

& TReconLM
& 48.19
& 98.72
& 1.3829
& 4.87 \\

& \textbf{CCP-DP}
& \textbf{95.36}
& \textbf{99.89}
& \textbf{0.1221}
& 2.200 \\

\bottomrule
\end{tabularx}
\end{table}

\subsection{Multi-dataset ablation analysis}
\label{sec:supp_ablation}

We next examined how the major components of CCP-DP contribute to the
overall reconstruction process across Srinivas-110, Chandak-108, and
Bar-Lev-140. Rather than producing uniform gains on every dataset, the
different components play distinct roles: local guide construction provides
the initial sequence estimate, structured decoding resolves difficult
clusters, cross-cluster context supplies additional sequence evidence when
local support is insufficient, and adaptive routing avoids unnecessary
structured decoding.

The ablation conditions in Supplementary Table~\ref{tab:supp_ablation}
use different decoding scopes and scoring components.
``Provisional guide only'' directly outputs the first-stage provisional guide
without subsequent structured decoding.
``w/o Cross-LOO score'' retains adaptive routing and the remaining decoder
components but removes the Cross-LOO contribution from trellis scoring.
``w/o blind IDS likelihood'' retains blind IDS estimation but removes the
IDS likelihood term from trellis scoring.
``Full, decode-all'' retains the complete scoring scheme but disables
adaptive routing and applies structured decoding to every cluster.
``Local trellis, decode-all'' also decodes every cluster but removes both
the Cross-LOO score and blind IDS likelihood, leaving only local structured
evidence in the trellis.
``Full CCP-DP'' uses the complete scoring scheme together with adaptive
routing.

The Provisional guide only results illustrate the limitation of relying exclusively on
local cluster evidence. Exact reconstruction reached 92.86\% on
Srinivas-110 and 86.49\% on Chandak-108, compared with 98.01\% and
95.36\%, respectively, for the full method. Introducing structured decoding
therefore accounts for a substantial part of the recovery on these more
challenging datasets. The Local trellis, decode-all variant further improves over
Provisional guide only, but remains below the full method, indicating
that structured local inference alone does not recover the full benefit of
CCP-DP.

The clearest additional gain comes from the Cross-LOO contextual score.
When this term is removed, exact reconstruction decreases from 98.01\% to
94.14\% on Srinivas-110 and from 95.36\% to 91.41\% on Chandak-108.
The much smaller change on Bar-Lev-140 is consistent with the high accuracy
already achieved by the provisional guide on this dataset. These results
indicate that cross-cluster context is particularly valuable when local
evidence alone does not fully resolve the remaining reconstruction ambiguity.

The blind IDS likelihood produced comparatively modest changes in exact
reconstruction, while providing an explicit dataset-adaptive model of
substitution, insertion, and deletion tendencies. It therefore serves as a
complementary source of evidence alongside Cross-LOO.

The role of adaptive routing is most evident in computational efficiency.
On Srinivas-110, decoding all clusters yields the same exact reconstruction
rate as the full method (98.01\%), but increases runtime from 1.782~s to
6.824~s. On Bar-Lev-140, the corresponding runtime increases from 2.893~s
to 4.453~s, again with little change in reconstruction accuracy. Thus, adaptive
routing primarily concentrates structured decoding on the subset of clusters
for which it is most useful, avoiding unnecessary computation on
well-supported clusters.

Taken together, the ablation results support the intended division of labor
within CCP-DP: provisional guides provide an efficient local estimate,
structured decoding repairs difficult cases, Cross-LOO contributes additional
cross-cluster evidence on challenging datasets, and adaptive routing preserves
these gains while reducing unnecessary computation.

\begin{table}[t]
\centering
\caption{
\textbf{Multi-dataset ablation analysis of CCP-DP.}
Exact reconstruction rates and runtimes are reported for three real datasets
under different component configurations. ``Full, decode-all'' retains the
complete scoring scheme but disables adaptive routing, whereas ``Local
trellis, decode-all'' applies structured decoding to every cluster using only
local trellis evidence.
}
\label{tab:supp_ablation}

\small
\setlength{\tabcolsep}{3pt}
\renewcommand{\arraystretch}{1.12}
\renewcommand{\tabularxcolumn}[1]{m{#1}}

\begin{tabularx}{\textwidth}{@{}>{\hsize=1.4\hsize\raggedright\arraybackslash}X*{6}{>{\hsize=0.9333\hsize\centering\arraybackslash}X}@{}}
\toprule
\multirow{2}{*}{\textbf{Variant}} &
\multicolumn{2}{c}{\textbf{Srinivas-110}} &
\multicolumn{2}{c}{\textbf{Chandak-108}} &
\multicolumn{2}{c}{\textbf{Bar-Lev-140}} \\
\cmidrule(lr){2-3}
\cmidrule(lr){4-5}
\cmidrule(lr){6-7}

&
\textbf{Exact (\%)} &
\textbf{Time (s)} &
\textbf{Exact (\%)} &
\textbf{Time (s)} &
\textbf{Exact (\%)} &
\textbf{Time (s)} \\
\midrule

Provisional guide only
& 92.86 & 1.222
& 86.49 & 0.883
& 98.41 & 1.181 \\

w/o Cross-LOO score
& 94.14 & 1.615
& 91.41 & 1.923
& 98.74 & 2.419 \\

w/o blind IDS likelihood
& 97.94 & 1.766
& 95.50 & 2.133
& 98.75 & 2.322 \\

Full, decode-all
& 98.01 & 6.824
& 95.36 & 2.230
& 98.76 & 4.453 \\

Local trellis, decode-all
& 94.86 & 5.145
& 92.36 & 1.863
& 98.77 & 4.342 \\

\textbf{Full CCP-DP}
& \textbf{98.01} & 1.782
& \textbf{95.36} & 2.200
& \textbf{98.74} & 2.893 \\

\bottomrule
\end{tabularx}
\end{table}

\subsection{Diagnostic fixed-weight analysis of Cross-LOO context}
\label{sec:supp_fixed_crossloo}

To assess whether Cross-LOO context remains informative on datasets where
its contribution is rejected by the standard calibration procedure, we
performed an additional diagnostic analysis on
Bar-Lev-140~\citep{barlev2025scalable},
Grass-117~\citep{grass2015robust}, and
Welter-110 subset (file 2)~\citep{welter2026end}.
Under standard CCP-DP, the effective Cross-LOO weight
$\lambda_c^{\mathrm{eff}}=\tau\lambda_c^0$ was reduced to zero on all
three datasets by the trust-calibration procedure.

For this diagnostic analysis, we bypassed the calibration outcome and
directly fixed the effective Cross-LOO weight used in trellis scoring to
$w_{\mathrm{cross}}=\lambda_c^{\mathrm{eff}}=0.50$.
We then compared No Cross-LOO, Correct Cross-LOO, Table-permuted, and
Wrong-dataset context, while keeping all remaining decoder settings
unchanged.

As shown in Supplementary Table~\ref{tab:supp_fixed_crossloo},
Correct Cross-LOO consistently outperformed both Table-permuted and
Wrong-dataset context, indicating that the matched context tables retain
dataset-specific information. However, relative to No Cross-LOO, the effect
was dataset dependent: exact reconstruction changed by $-0.610$, $+0.020$,
and $-7.570$ percentage points on Bar-Lev-140, Grass-117, and
Welter-110 subset (file 2), respectively. These results indicate that
informative cross-cluster structure does not necessarily improve
reconstruction when applied with a fixed weight, and support the use of
dataset-level trust calibration to suppress contextual evidence when its
predictive contribution is insufficient.

\begin{table}[t]
\centering
\caption{
\textbf{Diagnostic fixed-weight analysis of Cross-LOO context.}
For Correct Cross-LOO, Table-permuted, and Wrong-dataset conditions,
the standard calibration outcome was bypassed and the effective trellis
weight was fixed at
$w_{\mathrm{cross}}=\lambda_c^{\mathrm{eff}}=0.50$.
No Cross-LOO instead set the contextual contribution to zero.
All clusters were decoded under otherwise identical settings.
$\Delta$Exact denotes the change in exact reconstruction of Correct
Cross-LOO relative to No Cross-LOO.
}
\label{tab:supp_fixed_crossloo}

\small
\setlength{\tabcolsep}{3pt}
\renewcommand{\arraystretch}{1.12}
\renewcommand{\tabularxcolumn}[1]{m{#1}}

\begin{tabularx}{0.90\textwidth}{@{}>{\hsize=1.3\hsize\centering\arraybackslash}X*{5}{>{\hsize=0.94\hsize\centering\arraybackslash}X}@{}}
\toprule
\textbf{Dataset} &
\shortstack{\textbf{No} \\ \textbf{Cross-LOO}} &
\shortstack{\textbf{Table-} \\ \textbf{permuted}} &
\shortstack{\textbf{Wrong-} \\ \textbf{dataset}} &
\shortstack{\textbf{Correct} \\ \textbf{Cross-LOO}} &
\shortstack{\boldmath$\Delta$\textbf{Exact} \\ \textbf{(pp)}} \\
\midrule

Bar-Lev-140
& 98.760
& 84.192
& 92.600
& 98.150
& $-0.610$ \\

Grass-117
& 99.458
& 96.584
& 98.776
& 99.478
& $+0.020$ \\

Welter-110 subset (file 2)
& 93.410
& 30.752
& 46.350
& 85.840
& $-7.570$ \\

\bottomrule
\end{tabularx}
\end{table}

\subsection{Parameter sensitivity analysis}
\label{sec:supp_sensitivity}

To assess the robustness of CCP-DP to parameter choices, we performed
an extended parameter-sensitivity analysis on Srinivas-110 and Chandak-108.
Six parameters associated with provisional-consensus construction, contextual
evidence, and adaptive routing were evaluated: the coverage switch, reuse
threshold, phase-load thresholds, Cross-LOO weight scale, minimum routing
fraction, and phase-band width. Each parameter was varied individually around
its default value while all remaining settings were held fixed. For all four
reported metrics, changes are measured relative to the corresponding default
configuration.

As shown in Supplementary Fig.~\ref{fig:supp_parameter_sensitivity_full},
reconstruction quality was stable over broad ranges of most parameters.
The reuse threshold, phase-load thresholds, and phase-band width produced
little or no change in either exact reconstruction rate or mean edit distance
over the tested ranges. The coverage switch and Cross-LOO weight scale showed
the clearest effects on reconstruction quality, particularly on Chandak-108.
For the coverage switch, performance degraded when the threshold was reduced
substantially below the default value of 40 reads, consistent with excessive
assignment of lower-coverage clusters to the high-coverage provisional-guide
branch. Similarly, the default Cross-LOO weight scale of 1.0 lay near the
best-performing region on both datasets, whereas larger deviations produced
moderate losses in reconstruction accuracy.

Computational behavior was more sensitive than reconstruction quality to
several routing-related parameters. Runtime changed more noticeably with the
phase-load thresholds, Cross-LOO weight scale, minimum routing fraction, and
phase-band width, particularly on Srinivas-110. In contrast, the fraction of
clusters entering structured decoding remained largely unchanged across most
parameter variations. The main exception was the minimum routing fraction on
Srinivas-110, for which increasing the prescribed minimum produced the
expected increase in the proportion of clusters routed to structured decoding.
Taken together, these results indicate that CCP-DP maintains stable
reconstruction quality over a reasonably broad neighborhood of its default
configuration, while several parameters primarily influence computational
allocation rather than reconstruction accuracy.

\begin{figure}[H]
    \centering
    \includegraphics[width=0.90\textwidth]
    {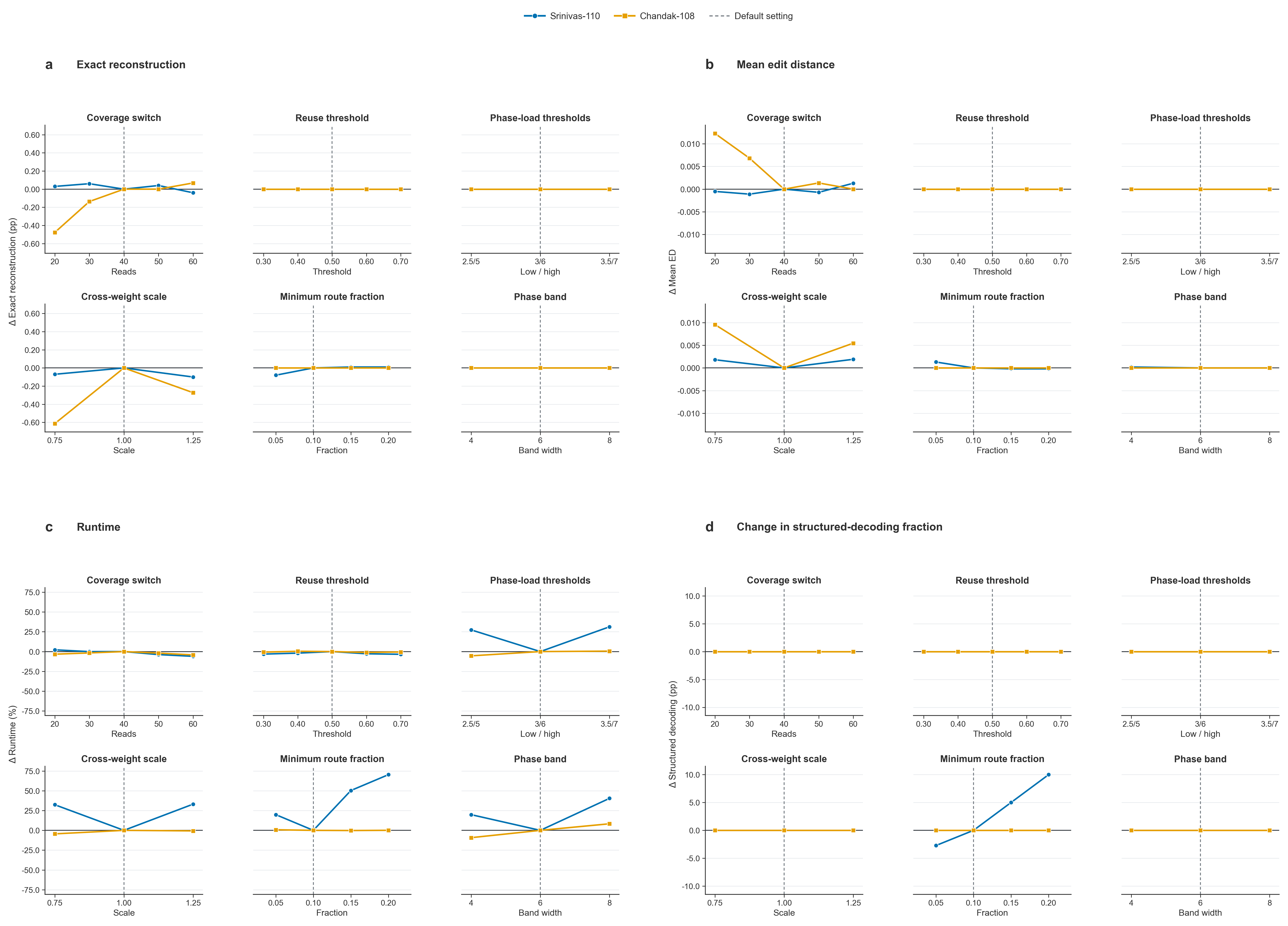}
    \caption{
    \textbf{Extended parameter-sensitivity analysis of CCP-DP.}
    Changes relative to the corresponding default configuration in
    \textbf{(a)} exact reconstruction rate,
    \textbf{(b)} mean edit distance,
    \textbf{(c)} runtime, and
    \textbf{(d)} the fraction of clusters routed to structured decoding
    for Srinivas-110 and Chandak-108.
    Within each metric block, six parameters are varied independently:
    the coverage switch, reuse threshold, phase-load thresholds,
    Cross-LOO weight scale, minimum routing fraction, and phase-band width.
    All remaining parameters are held at their default settings.
    Horizontal reference lines indicate no change relative to the default
    configuration, whereas vertical dashed lines indicate the corresponding
    default parameter values.
    The six subpanels within each metric block share a common vertical scale
    to permit direct comparison of sensitivity across parameters.
    Positive values in \textbf{(a)} indicate improved exact reconstruction;
    negative values in \textbf{(b)} and \textbf{(c)} indicate lower edit
    distance and shorter runtime, respectively.
    Positive values in \textbf{(d)} indicate an increased fraction of
    clusters routed to structured decoding relative to the default setting.
    }
    \label{fig:supp_parameter_sensitivity_full}
\end{figure}

\subsection{Mean edit-distance robustness across sequencing coverages}
\label{sec:supp_coverage_mean_ed}

To complement the exact-reconstruction analysis, we further evaluated
the mean edit distance under the same coverage conditions. The analysis
used the same sampled read clusters and random seeds as the corresponding
exact-reconstruction experiments.

As shown in Supplementary Fig.~\ref{fig:supp_coverage_mean_ed}, mean edit
distance generally decreases as sequencing coverage increases. These results
are consistent with the exact-reconstruction trends and further indicate that
CCP-DP maintains low residual reconstruction error across a broad range of
coverage conditions, including the low-coverage regime.

\begin{figure}[H]
    \centering
    \includegraphics[width=0.78\textwidth]
    {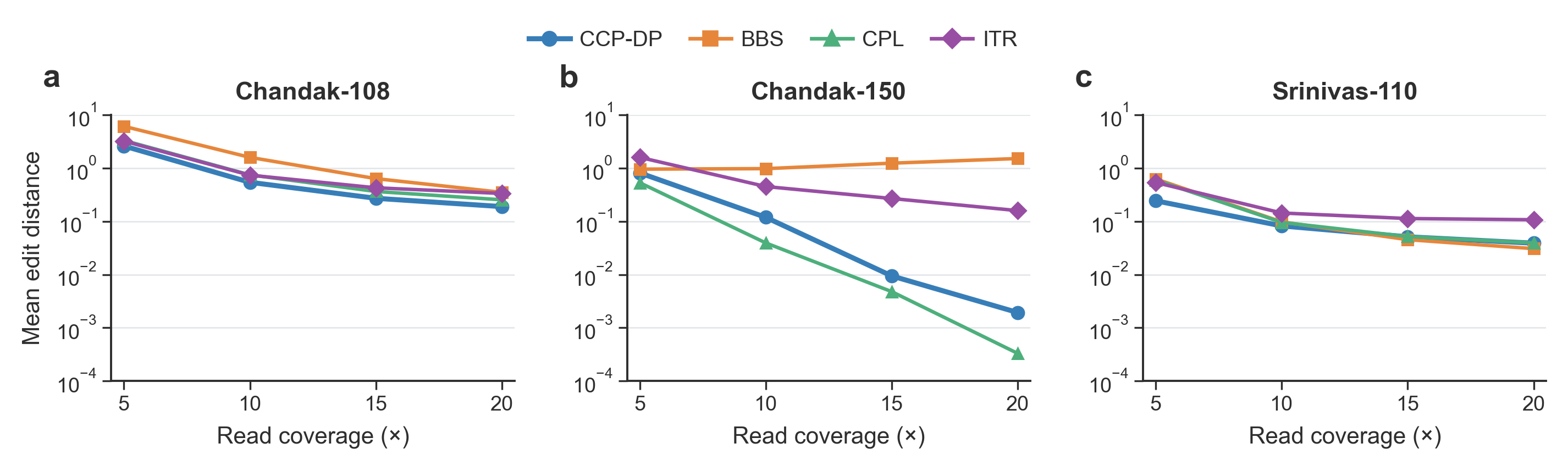}
    \caption{
    \textbf{Mean edit distance across sequencing coverages.}
    Mean edit distances from \(5\times\) to \(20\times\) coverage are shown
    for \textbf{(a)} Chandak-108, \textbf{(b)} Chandak-150, and
    \textbf{(c)} Srinivas-110. CCP-DP is compared with BBS, CPL, and ITR
    using the same sampled read clusters as in the corresponding
    exact-reconstruction experiment. Points show the mean over five random
    seeds, and error bars indicate standard deviations.
    }
    \label{fig:supp_coverage_mean_ed}
\end{figure}

\end{document}